\documentclass[onecolumn,showpacs,nofootinbib,11pt]{revtex4-1}

\usepackage{color}
\usepackage{graphicx}
\usepackage{graphicx,float}\usepackage[all]{xy}
\usepackage{amsmath,upgreek}

\usepackage{amssymb}

\newcommand{\beq}{\begin{eqnarray}}
\newcommand{\eeq}{\end{eqnarray}}
\newcommand{\be}{\begin{equation}}
\newcommand{\ee}{\end{equation}}
\newcommand{\mr}{\mathring}
\newcommand{\bea}{\begin{eqnarray}}
\newcommand{\eea}{\end{eqnarray}}

\newcommand{\ba}{\begin{eqnarray}}
\newcommand{\ea}{\end{eqnarray}}
\usepackage[colorlinks,hyperindex]{hyperref}
\definecolor{green1}{RGB}{0,128,0} 
\hypersetup{hidelinks,backref=true,pagebackref=true,hyperindex=true,colorlinks=true,breaklinks=true,urlcolor= blue}
\hypersetup{%
  colorlinks = true,
  linkcolor  = blue,
  citecolor = green1,
}
\usepackage{bookmark,textgreek}
\usepackage{hyperref,color,xcolor}
\hypersetup{hidelinks,hyperindex=true,colorlinks=true,breaklinks=true,urlcolor= blue}
\hypersetup{%
  colorlinks = true,
  linkcolor  = blue
}
\begin{document}
\title{MGD Dirac stars}

\author{Rold\~ao da Rocha}
\affiliation{CMCC,  Federal University of ABC, 09210-580, Santo Andr\'e, Brazil.}
\email{roldao.rocha@ufabc.edu.br}

%
%


\begin{abstract} 
The method of geometric deformation (MGD) is here employed to study  compact stellar configurations, which are solutions of the effective Einstein--Dirac coupled field equations on fluid branes. Non-linear, self-interacting, fermionic fields are then employed to derive MGD Dirac stars, whose properties are analyzed and  discussed.  The MGD Dirac star maximal mass is shown to increase as an specific function of the spinor self-interaction coupling constant, in a realistic model 
involving the most strict phenomenological current bounds for the brane tension.

\end{abstract}



\maketitle

\section{Introduction} 

The method of geometrical deformation (MGD) is a protocol 
emerging in the context of the AdS/CFT membrane paradigm, having general relativity (GR) as a limit that corresponds to rigid branes   \cite{Ovalle:2017fgl,ovalle2007,Casadio:2012rf,darkstars}. The MGD well describes compact stellar configurations in braneworld scenarios  \cite{Antoniadis:1998ig,Antoniadis:1990ew,Randall:1999ee}.  The MGD has the brane tension, $\upsigma$, as a running parameter that encrypts Kaluza--Klein modes and gravity in the bulk as well.  A rigid brane corresponds to the GR, $\upsigma\to\infty$, limit. As the brane emulates the universe we live in, cosmological  expansion can be implemented from the brane deformation, due to a finite brane tension, into the  warped additional dimension of the bulk. In fact, 
realistic brane-world models take into account a variable brane tension, that is proportional to the brane temperature, according to the E\"otv\"os law \cite{Abdalla:2009pg,daRocha:2012pt,Gergely:2008jr}.

The MGD is a procedure that derives solutions of the Einstein field equations on the brane, encompassing compact stellar  configurations \cite{covalle2,darkstars,Ovalle:2007bn,ovalle2007}. 
As a natural step in the membrane paradigm of AdS/CFT \cite{BoschiFilho:2001gr,BoschiFilho:2003zi}, the    holographic entanglement entropy of MGD black holes was computed in Ref.  \cite{daRocha:2019pla}. 
 The MGD has observational bounds, derived in Refs. \cite{Cavalcanti:2016mbe,Casadio:2015jva,Casadio:2016aum,Fernandes-Silva:2019fez} in several physical contexts. 
 In the MGD and its extensions (EMGD) \cite{Casadio:2013uma,Casadio:2015gea,Ovalle:2013vna,Ovalle:2019qyi,Ovalle:2013xla}, the finite brane tension governs the deformation of the Schwarzschild solution, in such a way that brane/bulk  effects can be  implemented.  The information entropy was employed to predict critical densities of several types of stars, in the MGD and EMGD contexts \cite{Casadio:2016aum,Fernandes-Silva:2019fez}. Besides, MGD dark glueball stars implementing hidden SU($N$) gauge theories \cite{Bernardini:2016qit}, whose signature may be detected in eLISA, were scrutinized in Ref. \cite{daRocha:2017cxu}. EMGD compact glueball stellar configurations were also studied in Ref. \cite{Fernandes-Silva:2018abr}. Hydrodynamical analogues of MGD black holes were discussed in Ref. \cite{daRocha:2017lqj}. The MGD was also used in Refs. \cite{Contreras:2018gzd,Casadio:2019usg,Rincon:2019jal,Ovalle:2019lbs,Gabbanelli:2019txr,Contreras:2019mhf,Ovalle:2018ans,Morales:2018urp,Morales:2018nmq,Panotopoulos:2018law}, that   include anisotropic solutions of quasi-Einstein equations  \cite{Ovalle:2017wqi,Ovalle:2017fgl,Gabbanelli:2018bhs,PerezGraterol:2018eut,Heras:2018cpz}.  Besides, the MGD was employed in the context of the generalized uncertainty principle to study the MGD black hole thermal spectrum \cite{Casadio:2017sze}.  

The so called Dirac stars encompass compact stellar distributions generated by fermionic fields~\cite{Herdeiro:2017fhv,Dzhunushaliev:2018jhj}. 
 Exact solutions for self-gravitating Dirac
                        fields were studied in Ref. \cite{Cianci:2016pvd}.
Refs. \cite{Fabbri:2014zya,Fabbri:2014wda,Fabbri:2011mi} also scrutinize fermionic fields in gravity. Besides,  Ref.~\cite{Dzhunushaliev:2019uft} discussed Dirac stars generated by non-Abelian gauge fields. The main aim of this work is to study MGD compact stellar configurations, that are solutions of the effective Einstein--Dirac coupled field equations.
The MGD-decoupling method is employed to correct the effective energy-momentum tensor on the brane by fermionic background effects. Non-linear, self-interacting, massive spinor fields are used to derive physical properties of MGD Dirac stellar configurations, with respect to both the finite brane tension and the spinor self-interaction coupling constant. The MGD Dirac star stability is also discussed.

This paper is organized as follows: Sect. \ref{MGD} is devoted to introduce a brief review, regarding the MGD applied to stellar distributions on a fluid brane, ruled by a variable tension that encodes the cosmological evolution. In Sect. \ref{dsfb} the Einstein--Dirac coupled system of ODEs is solved for self-interacting fermionic fields, having the MGD as a natural input. Then, the solutions of the coupled system of ODEs represent MGD Dirac stars, whose observational features are scrutinized and illustrated in several regimes. In Sect. \ref{4} the conclusions and perspectives are presented.

\section{The MGD setup and fluid branes}
\label{MGD}
 The MGD procedure is constructed to derive high energy corrections to GR \cite{ovalle2007,Ovalle:2013xla,darkstars}. 
Fluid branes have a variable tension that emulates cosmological evolution  \cite{Gergely:2008jr,Abdalla:2009pg}. 
The extended MGD has been recently employed to derive, in the context of the quantum portrait of black holes, the strictest brane tension bound $\upsigma \gtrsim  2.81\times10^6 \;{\rm MeV^4}$ \cite{Fernandes-Silva:2019fez}. 

The brane Einstein field equations can be derived, by the Gauss-Codazzi projection method, from the Einstein equations in the bulk. Hereon $c=8\pi G_4=1$, for $G_4 = \frac{\hbar c}{M_{\rm p}^2}$ is the brane coupling constant, where $M_{\rm p}$ denotes the Planck mass.
The brane metric is governed by the brane effective Einstein field equations
\begin{equation}
\label{5d4d}
G_{\mu\nu}
=\Uplambda g_{\mu\nu}+\mathbb{T}_{\mu\nu},
\end{equation} where $G_{\mu\nu}=R_{\mu\nu}-\frac12 Rg_{\mu\nu}$ is the Einstein tensor and $\Uplambda$ stands for the cosmological running parameter on the brane. 
The energy-momentum tensor in (\ref{5d4d}) can be split into the following components, 
\beq
\mathbb{T}_{\mu\nu}
=
T_{\mu\nu}+\mathbb{E}_{\mu\nu}+\frac{1}\upsigma S_{\mu\nu}
+L_{\mu\nu}+P_{\mu\nu}.\label{tmunu}\eeq 
The term $T_{\mu\nu}$ denotes the energy-momentum encoding all kind of matter on the brane and $\mathbb{E}_{\mu\nu}$ is the electric component of the bulk  Weyl curvature tensor, taking account the average over the two edges of the $\mathbb{Z}_2$ symmetric brane, whereas   
$S_{\mu\nu}$ is a tensor whose intricate expression involves quadratic terms of energy-momentum of the brane. Besides, the term $L_{\mu\nu}$ comes from an eventual asymmetric embedding of the brane into the bulk, whereas $P_{\mu\nu}$ regards the pull back of  fields in the bulk that are beyond the standard
model, encompassing moduli fields, dilatons and quantum radiation, for instance \cite{Gergely:2008jr}.

Stellar configurations that can be modeled by solutions of Eq. (\ref{5d4d}) are static and spherically symmetric, described by a metric of type \begin{equation}\label{abr}
ds^{2} = -A(r) dt^{2} + \frac{dr^{2}}{B(r)} + r^{2} (d\uptheta^2+\sin^2\uptheta d\upphi^2). 
\end{equation} 
The MGD metric was initially derived in Refs.  \cite{covalle2,darkstars,Ovalle:2007bn},  reading 
\begin{eqnarray}
\label{deff}
B(r)
=
{1-\frac{2\,M}{r}}
+\upsigma\,e^{
\int^r_{{\rm R}}d r'\,
{f(r')}/{g(r')}\,}
\ ,
\label{I}
\end{eqnarray}
for a stellar configuration of mass $M$, where \cite{covalle2} \beq
f(r)&=&\frac{1}{r^{2}}\frac{A(r)A''(r)}{A'^{2}(r)}\!+\!(\log[A(r)])^{\prime\,2}-1+\!\frac{2}{r}\log[A(r)]^{\prime}\!\\ g(r)&=&\frac{1}{\frac12\log[A(r)]^{\prime}+\frac{2}{r}},\eeq and  \beq
{R} = \frac{\int\,dr\, r^3\,\uprho(r)}{\int\,dr\, r^2 \uprho(r)}\eeq is the effective stellar configuration radius, for $\uprho(r)$ being the
compact stellar configuration energy density \cite{ovalle2007}. The brane tension $\upsigma$ in Eq. (\ref{deff}) governs 
the influence of bulk effects onto the brane, carrying information about the bulk Weyl fluid bathing the brane  \cite{Casadio:2015jva}. After neglecting  the 
infinitesimal values of higher order powers of the small observational brane tension, the MGD metric can be rewritten as  \cite{ovalle2007,covalle2}
\begin{subequations}
\ba
\label{nu}
\!\!\!\!\!\!A(r)
&=&
1-\frac{2\,M}{r}
\ ,
\\
\!\!\!\!\!\!B(r)
&=&
\left[1+\frac{l{\upsigma}}{r-\frac{3\,M}{2}}\,\right]\left(1-\frac{2\,M}{r}\right)
\ ,
\label{mu}
\ea
\end{subequations} 
where 
\begin{equation}
\label{L}
l
=
\left(R-\frac{3M}{2{}}\right)\left(1-\frac{2M}{{R}}\right)^{\!-1}\!{}.
\end{equation} 
In  the general-relativistic 
limit $\upsigma\to\infty$, the Schwarzschild metric is recovered from the MGD metric.

\section{Dirac stars on fluid branes}
\label{dsfb}

Compact gravitating configurations on fluid branes, modeled by the MGD 
with a background spinor field, $\psi$, of mass $m$,  
can be described by the Einstein field equations (\ref{5d4d}), coupled to a Dirac equation.  To the action that generates the Einstein field equations, it must be added the spinor  Lagrangian \begin{equation}
	\mathcal{L}_\psi =	\frac{i \hbar}{2} \left(
			\bar \psi \gamma^\mu \nabla_\mu\psi -
			\bar \psi \overleftarrow{\nabla}_{\mu} \gamma^\mu \psi
		\right) - m \bar \psi \psi - \Omega_\psi,
\label{lgian}
\end{equation}
where self-interacting spinor fields are implemented by  
\begin{equation}
	\Omega_\psi = - \frac{\uplambda}{2} \left(\bar\psi\psi\right)^2.
\label{selfii}
\end{equation}
Here $\{\gamma^\mu\}$ is the set of gamma matrices, that satisfies the Clifford algebra $\{\gamma^\mu, \gamma^\nu\}=2g_{\mu\nu}\mathbb{I}_{4\times 4}$, and $\bar\psi=\psi^\dagger\gamma^0$ is the spinor conjugate. 
Eq. (\ref{selfii}) displays non-linear, self-interacting, spinor fields. In fact, compact stellar configurations, which are solutions of the Einstein--Dirac coupled system of equations for  \emph{linear} spinor fields, have mass much smaller than the Chandrasekhar mass. Ref. \cite{Dzhunushaliev:2018jhj} considered self-interacting spinor fields, demonstrating that the mass of compact stellar configurations increase several orders of magnitude, encompassing astrophysical stars. A similar approach will be used hereon, in the MGD setup.

 Eq. (\ref{5d4d}) can be obtained by varying the bulk Einstein--Hilbert action, when one uses the Gauss--Codazzi equations. In the presence of a spinor field background, one derives the Einstein field equations and the Dirac equation, using the MGD-decoupling method \cite{Ovalle:2017fgl}  
\begin{eqnarray}
	G_{\mu\nu} &=&
	\Uplambda g_{\mu\nu}+ \mathbb{T}_{\mu\nu} + (1+\zeta)\mr{T}_{\mu\nu},
\label{eins} \\
	i \hbar \gamma^\mu \nabla_\mu\psi - \uplambda  \psi - \frac{\partial \Omega_\psi}{\partial\bar\psi}&=& 0,
\label{einssp}\\
	i \hbar \bar\psi \overleftarrow{\nabla}_{\mu}  \gamma^\mu + \uplambda \bar\psi +
	\frac{\partial \Omega_\psi}{\partial\psi}&=& 0,
\label{einssp1}
\end{eqnarray} where  $\zeta$ is a small parameter driving the 
 MGD decoupling, and the fermionic energy-momentum tensor reads 
\begin{equation}
	\mr{T}_{\mu\nu} =\frac{i\hbar}{4}g_{\nu}^{\;\,\rho}\left(\bar\psi \gamma_{(\mu} \nabla_{\rho)}\psi+\bar\psi \overleftarrow{\nabla}_{(\mu}\gamma_{\rho)}\psi
\right)-\delta_{\mu\nu} \mathcal{G}_\psi,
\label{stensor}
\end{equation} for $\mathcal{G}_\psi=-\Omega_\psi+\frac12\left(\bar\psi\frac{\partial \Omega_\psi}{\partial\bar\psi}+\frac{\partial \Omega_\psi}{\partial\psi}\psi\right)$.
Eq. (\ref{eins}) represents the brane Einstein field equations (\ref{5d4d}), now corrected by the MGD-decoupling  
 for the spinor energy-momentum tensor $\mr{T}_{\mu\nu}$ in Eq. (\ref{stensor}).

In order to describe the background spinor field, the following two stationary ans\"atze  for $\psi$ are  
compatible with the metric \eqref{abr} \cite{Herdeiro:2017fhv,Dzhunushaliev:2018jhj}, given by 
\beq
\psi_1&=&\sqrt{2} e^{-i \frac{E t}{\hbar}} \left(\begin{matrix}
			0 \\ -ie^{i(\theta-\varphi)/2}\beta\\i \alpha \sin \theta e^{- i \varphi}\\- i \alpha \cos \theta\end{matrix}\right),\qquad\quad
	\psi_2=\sqrt{2} e^{-i \frac{E t}{\hbar}} \left(\begin{matrix}			- \beta \\0\\
		 \alpha \cos \theta e^{-i(\theta-\varphi)/2} \\  -\alpha \sin \theta e^{i(\theta-\varphi)/2} \\
		\end{matrix}\right),
	\label{spinor}
\eeq
where   $\alpha(r),\beta(r)$ are real functions.
When one replaces  \eqref{spinor}  into the field equations (\ref{eins}, \ref{einssp}), it yields the coupled systems of ODEs involving also the MGD metric (\ref{abr}):
\begin{eqnarray}
	&&\tilde \alpha^\prime + \left[
		\frac{A^\prime}{4\sqrt{AB}} + \frac{1}{x}\left(1+\frac{1}{\sqrt{B}}\right)
	\right] \tilde \alpha + \frac{1}{\sqrt{A}}\left(
		 1- \frac{\tilde E}{ \sqrt{A}} +
		8\tilde \uplambda({\tilde \alpha^2 - \tilde \beta^2})
	\right)\tilde \beta= 0,
\label{fe1}\\
	&&\tilde \beta^\prime + \left[
			\frac{A^\prime}{4 \sqrt{AB}} +
			\frac{1}{x}\left(1 - \frac{1}{\sqrt{A}}\right)
	\right]\tilde \beta  
		+ \frac{1}{\sqrt{B}}\left(
		 1- \frac{\tilde E}{ \sqrt{A}} +
		8\tilde \uplambda({\tilde \alpha^2 - \tilde \beta^2})
	\right)\tilde \alpha= 0,
\label{fe2}\\
	 &&\tilde m^\prime=8 x^2\left[
(\tilde \alpha^2 + \tilde \beta^2)\frac{\tilde E (1+\zeta)}{\sqrt{A}}+4\tilde \uplambda\left(\tilde \alpha^2 - \tilde \beta^2\right)^2
\right],
\label{fe3}
\end{eqnarray}
where, denoting $\uplambda_0=\hbar/m c$, the following quantities are defined \cite{Dzhunushaliev:2018jhj}:
\beq
\label{dmls_var}
	\tilde \alpha &=& 2\sqrt{\pi \uplambda_0^{3}}\frac{m \alpha}{M_{\rm p}}, \qquad \tilde \beta = 2\sqrt{\pi \uplambda_0^{3}}\frac{m\beta}{M_{\rm p}} , \qquad x = \frac{r}{\uplambda_0}, \qquad
	\tilde E = \frac{E}{m }, \qquad
	\\ 
	\widetilde{M} &=& \frac{m M}{M_{\rm p}^2}, \qquad
	\tilde \uplambda = \frac{\uplambda M_{\rm p}^2}{4\pi \uplambda_0^3m^3}.
\eeq
The mass of compact stellar configurations, 
$M\sim \frac{M_{\rm p}^2}{m}$, is usually studied when the spinor field mass, $m$, is much smaller than the Chandrasekhar mass, $\frac{M_{\rm p}^3}{m^2}$.

 As the MGD metric (\ref{abr}) is 
a solution of the Einstein field equations, we want now 
to verify which are the spinor fields components in (\ref{spinor}) that satisfy the coupled system (\ref{fe1}) -- (\ref{fe3}). Hence, numerical integration can be implemented, using boundary conditions near the stellar configuration center, as in Ref. \cite{Dzhunushaliev:2018jhj}:
\beq
	\tilde \beta= \tilde \beta_{\rm c} + \upbeta x^2+\ldots, \qquad
	\tilde \alpha=  \upalpha x+\ldots,
	 \qquad
	\tilde m=  {\rm m} x^3+\ldots,\label{exps}
\eeq
where quantities indexed by ``c'' stand for central ($x\approx 0$) values of the respective variables.
Some of expansion coefficients  in (\ref{exps}) can be derived once Eqs.~\eqref{fe1} - \eqref{fe3} are taken into account. On the other hand, the parameters $\tilde \beta_{\rm c}$ and $\tilde E$ 
are constrained just in order that the Minkowski spacetime be a solution of Eqs.~\eqref{fe1} - \eqref{fe3},  in the limit $r\to\infty$, $\zeta\to\infty$ and $\upsigma\to\infty$.

\begin{figure}[H]
\centering\includegraphics[width=9.6cm]{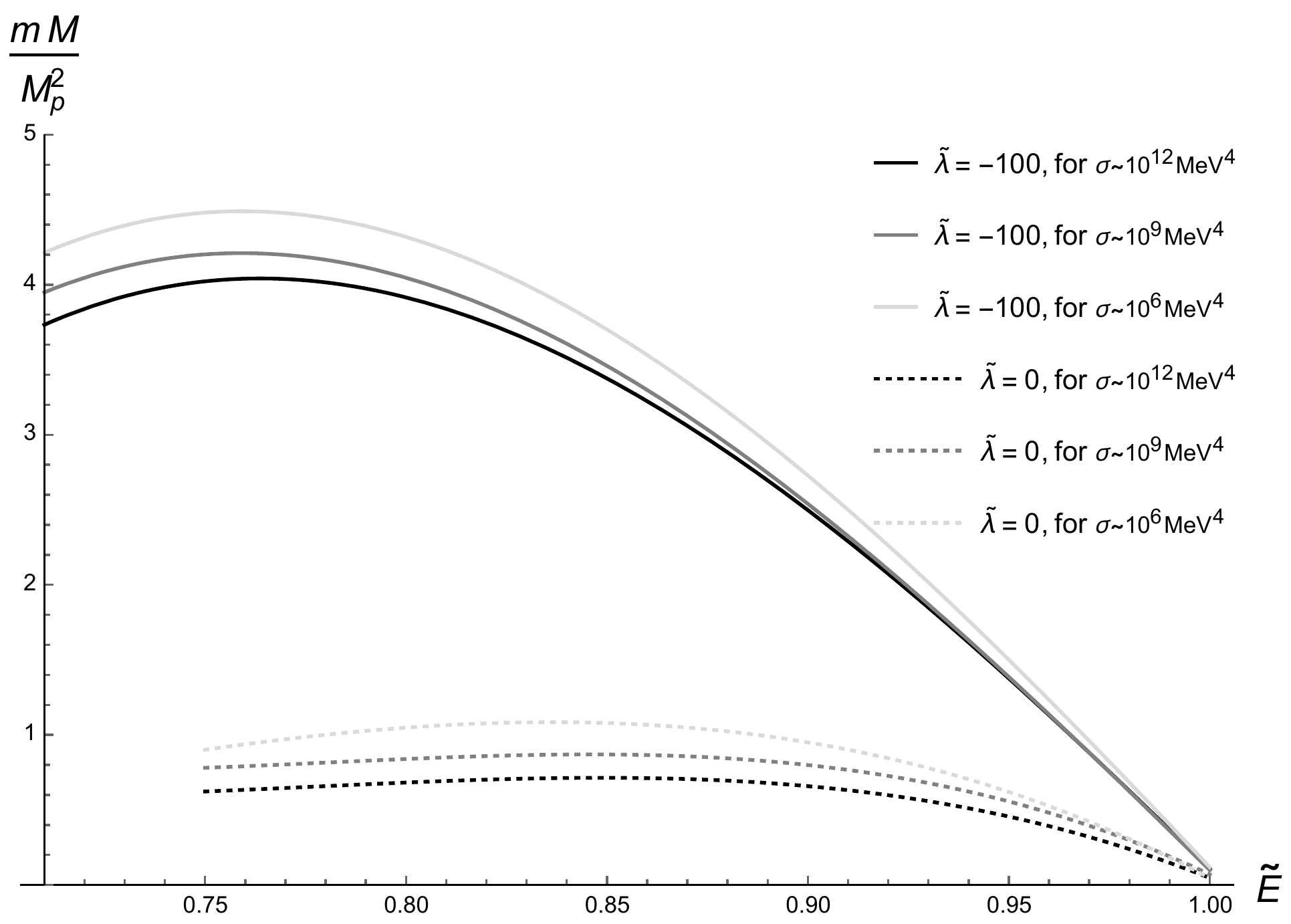}
\caption{MGD Dirac star mass, 
 as a function of $\tilde E$, shown for two values of the  coupling constant, $\tilde \uplambda$, and for $\upsigma \sim 3\times 10^6 \;{\rm MeV^4}$, $\upsigma \sim 10^9 \;{\rm MeV^4}$ and $\upsigma \sim 10^{12} \;{\rm MeV^4}$. The MGD-decoupling parameter $\zeta=0.1$ is adopted.}
\label{fi1}
\end{figure}

 The lower limit of integration of Eqs.~\eqref{fe1} - \eqref{fe3} is the center of the compact stellar configuration, where $\tilde \beta_{\rm c}$ is determined. 
 On the other hand, the upper limit of integration is the effective stellar radius where the functions
$\tilde \alpha$, $\tilde \alpha^\prime$, $\tilde \beta$ and $\tilde \beta^\prime$ vanish
 \cite{Dzhunushaliev:2018jhj}. 
After numerical integration, the MGD compact stellar configuration mass parameter, $\widetilde{M} = \frac{m M}{M_{\rm p}^2}$, as a function of $\tilde E$,  has the profile illustrated in Fig.~\ref{fi1}. Two different values of the coupling constant $\uplambda$,  driving the spinor field self-interaction in Eq. (\ref{selfii}), are considered, as well as three different values of the finite brane tension. In fact, as the most strict current bound on the variable brane tension is $\upsigma \gtrsim  2.81\times10^6 \;{\rm MeV^4}$ \cite{Fernandes-Silva:2019fez}, then in what follows 
we take the lower brane tension limit $\upsigma \sim 3\times 10^6 \;{\rm MeV^4}$, together with $\upsigma \sim 10^9 \;{\rm MeV^4}$ and $\upsigma \sim 10^{12} \;{\rm MeV^4}$, to 
study the physical differences among these cases. It is worth to emphasize that the general-relativistic limit corresponds to a rigid brane, making $\upsigma\to\infty$ and $\zeta\to\infty$. 

\noindent In Fig.~\ref{fi1}, for fixed values of the spinor coupling constant $\tilde \uplambda$, 
 there is a peak of the mass, at some value of
$\tilde E$. The bigger the brane tension, the smaller the maximal mass is, for both analyzed values of $\tilde \uplambda$. Boson stars were studied in a similar context, where a maximal mass was identified to a transition point, splitting stable and unstable compact stellar  configurations  \cite{Gleiser:1988ih}. This aspect was emulated and explored for MGD and EMGD compact stellar configurations, from the point of view of the information entropy, by Refs.  \cite{Casadio:2016aum,Fernandes-Silva:2019fez}.

It is worth to study compact stellar configurations in the regime $|\tilde \uplambda| \gg 1$. 
Ref.~\cite{Herdeiro:2017fhv} derived the maximal mass $0.7092\, \frac{M_{\rm p}^2}{m}$ of standard Dirac stars.
The spinor field mass,  $m$,  of the order $1~\text{GeV}$ yields the maximal  MGD stellar mass $M\sim 10^{11}~\text{Kg}$, corresponding to small mass stellar configurations. In fact, the Sun mass reads $M_\odot \approx 1.989\times 10^{30}$ Kg. Therefore it prevents  positive values
of $\tilde \uplambda$, as it implies a decrement of the maximal stellar mass. Studying astrophysical compact stellar configurations requires $\tilde \uplambda<0$. Fig.~\ref{fi1} shows that the more $|\tilde \uplambda|$ increases, the bigger the maximal MGD stellar masses are. In addition, for fixed values of $\tilde E$, the mass peaks are bigger, the lower the brane tension is.

\begin{figure}[H]
\centering\includegraphics[width=8cm]{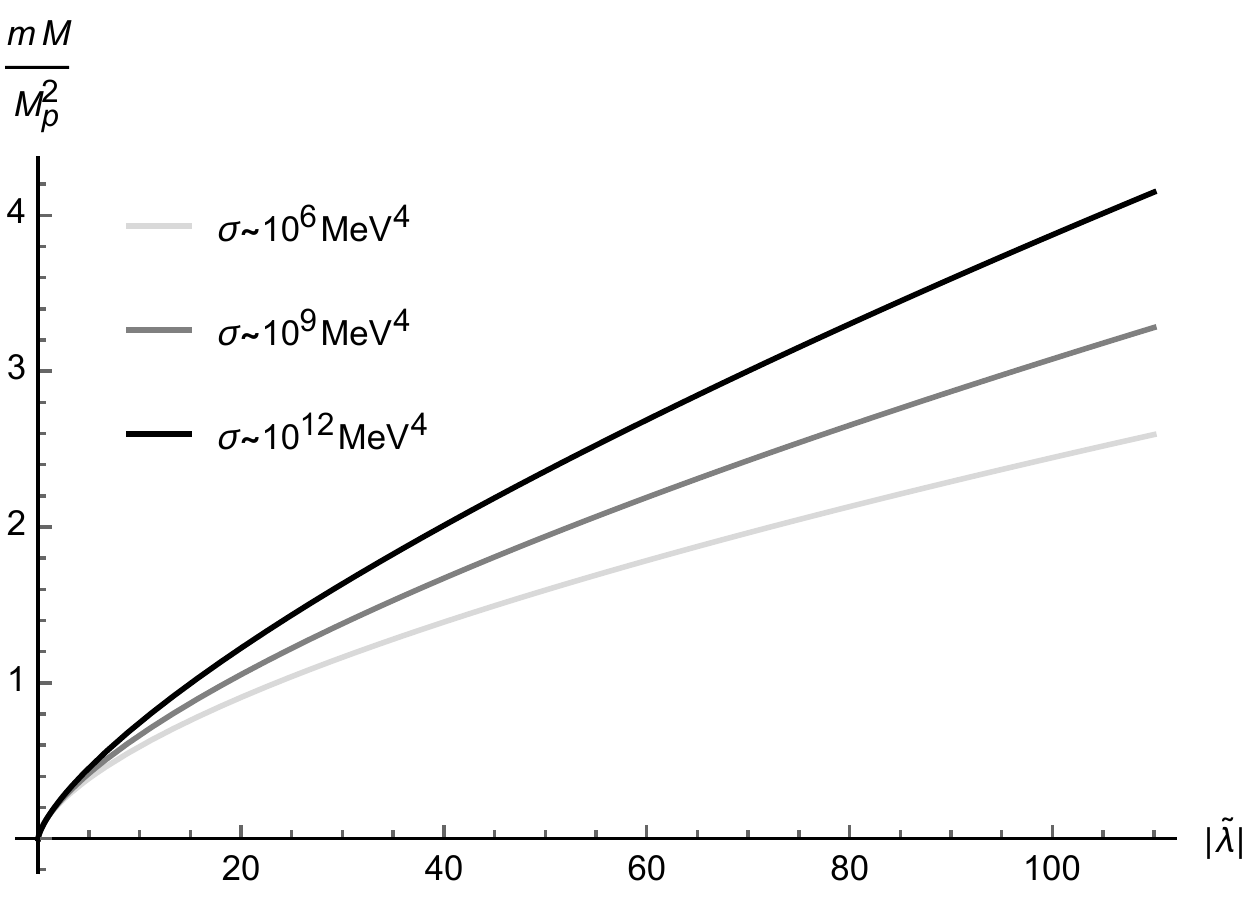}
\caption{MGD Dirac star maximal mass,
 $\tilde M$, as a function of $|\tilde \uplambda|$, for $\upsigma \sim 3\times 10^6 \;{\rm MeV^4}$, $\upsigma \sim 10^9 \;{\rm MeV^4}$ and $\upsigma \sim 10^{12} \;{\rm MeV^4}$. The MGD-decoupling parameter $\zeta=0.1$ is adopted. }
\label{fi2}
\end{figure}

Fig.~\ref{fi2} illustrates how the maximal MGD stellar mass
 increases as a function of $|\tilde \uplambda|$, for different values of the brane tension. 
\noindent Fig.~\ref{fi2} shows that the bigger the brane tension, the steeper the slope of each plot is. It indicates that more realistic models, involving observational values of the brane tension, yield a MGD Dirac star maximal mass that increases with $|\tilde \uplambda|$, however in a lower  rate. The general-relativistic limit $\upsigma\to\infty$, $\zeta\to\infty$ is acquired, being very close to the black line plot in Fig. \ref{fi2}, as indicated in Ref. \cite{Dzhunushaliev:2018jhj}. 
It is worth to emphasize that, when $\sigma\to\infty$, Fig. \ref{fi2} can be described, for $|\tilde \uplambda|\gg 1$, by  the interpolation expression 
 \begin{equation}
\label{41a}
	M^{\text{max}}= 0.4153 (1+\zeta)\sqrt{|\tilde \uplambda|}\frac{M_{\rm p}^2}{m}.
\end{equation} This emulates the results in Ref. \cite{Dzhunushaliev:2018jhj} in the MGD-decoupling context.

\begin{figure}[H]
\centering\includegraphics[width=8cm]{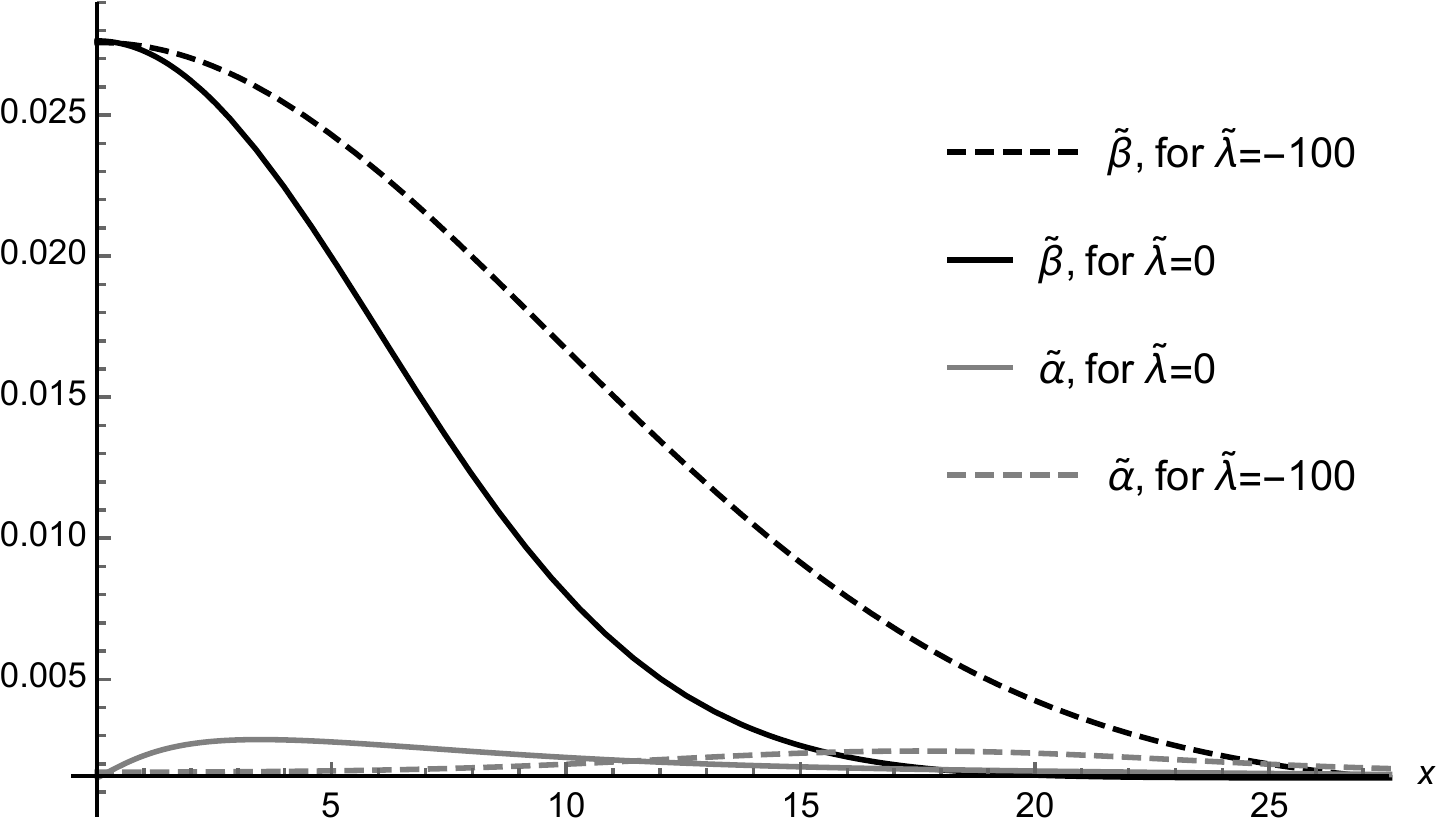}
\caption{Spinor fields profiles with respect to the Dirac star radius, $x$, for $\tilde \uplambda=0$ and $\tilde \uplambda=-100$, for a finite brane tension $\upsigma=10^{9} {\rm MeV}^4$. The MGD-decoupling parameter $\zeta=0.1$ is adopted.}
\label{fi3}
\end{figure}

When coupling spinor fields to the MGD solutions, 
the resulting compact stellar configurations present distinct profiles, in the $|\tilde \uplambda|\gg 1$ and the $|\tilde \uplambda|\approx0$ regimes.  Fig.~\ref{fi3} illustrates the spinor fields profiles with respect to the (adimensional radius) of the MGD Dirac star, for $\tilde \uplambda=0$ and $\tilde \uplambda=-100$.

\begin{figure}[H]
\centering\includegraphics[width=8cm]{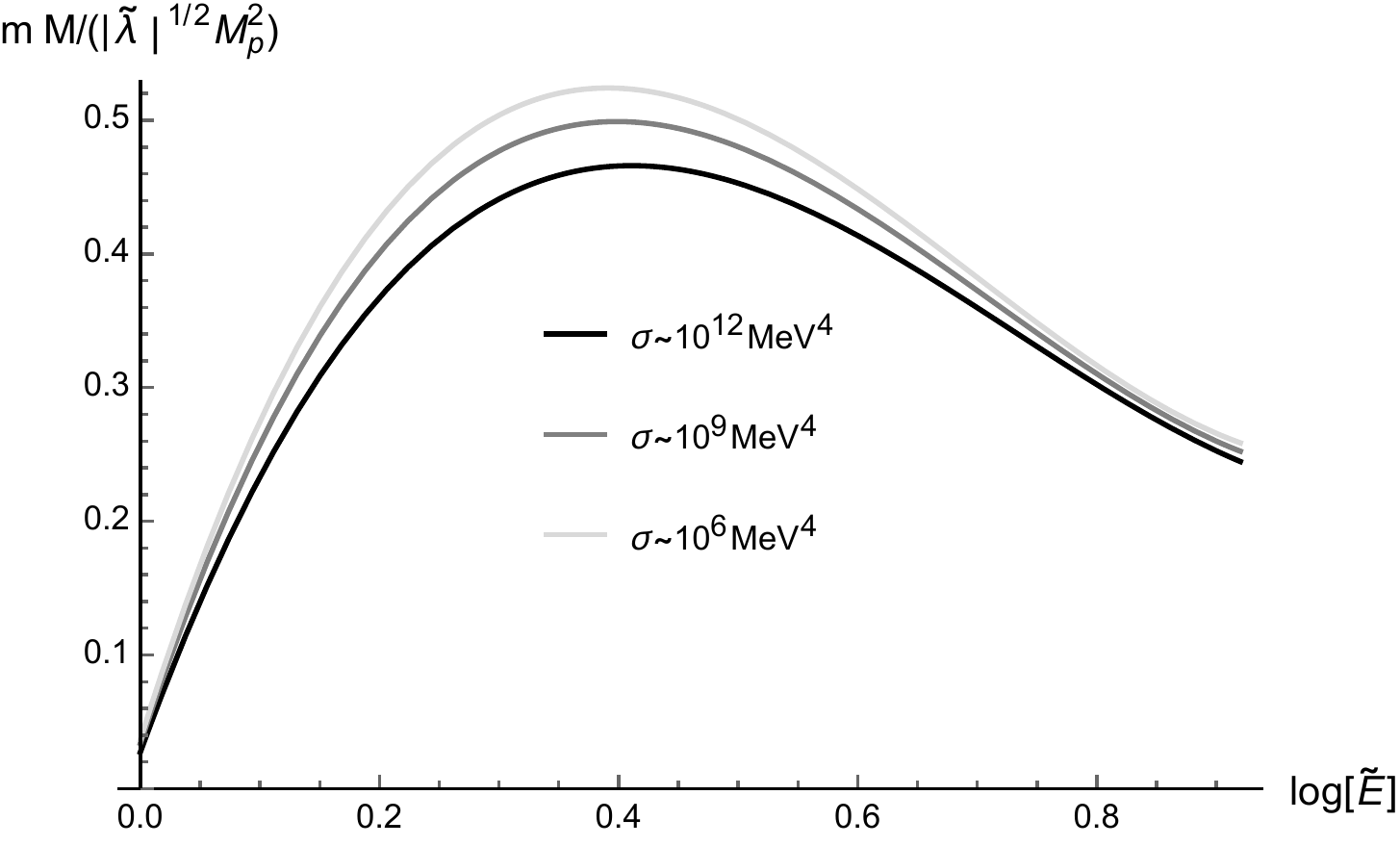}
\caption{Dimensionless MGD Dirac star mass $\tilde{M}_\star$, as a function
of $\tilde{E}$, for the limiting configurations described by  Eqs.~(\ref{ge0}, \ref{ge}), for $\upsigma \sim 3\times 10^6 \;{\rm MeV^4}$, $\upsigma \sim 10^9 \;{\rm MeV^4}$ and $\upsigma \sim 10^{12} \;{\rm MeV^4}$. The MGD-decoupling parameter $\zeta=0.1$ is adopted. }\label{fi4}
\end{figure}

Similar to bosonic stellar configurations \cite{Dzhunushaliev:2018jhj},  Eqs.~\eqref{fe1} - \eqref{fe3} can be made adimensional, at the 
$|\tilde \uplambda|\gg 1$ regime, by the mappings 
\begin{subequations}
\beq
\tilde \beta\mapsto\tilde \beta_\star &=& \sqrt{|\tilde \uplambda|}\tilde \beta,\qquad\quad\tilde \alpha\mapsto \tilde \alpha_\star=\sqrt{|\tilde \uplambda|}\tilde \alpha,\\
\tilde m\mapsto\tilde m_\star&=&\sqrt{|\tilde \uplambda|}\tilde m,\qquad\quad x\mapsto x_\star=\frac{x}{\sqrt{|\tilde \uplambda|}}.\eeq
\end{subequations}
Ref. \cite{Dzhunushaliev:2018jhj} showed that at a large $|\tilde \uplambda|$ regime, the fermionic field percolates a large range 
$\frac{\sqrt{|\tilde \uplambda|}}{m}$. This implies that terms involving $\tilde \alpha^\prime$ and
 $\tilde \beta^\prime$ can be disregarded. Hence, Eq. 
  \eqref{fe1} can be led to an analogue of Eq. \eqref{ge0}. 
Taking only leading terms in Eqs.~(\ref{fe1}, \ref{fe2})  yields, when $\tilde \beta\gg \tilde \alpha$, 
\begin{equation}
 \label{ge0}
 \tilde \beta_\star = \frac{1}{2\sqrt{2}}
  \left(\frac{\tilde E}{\sqrt{B}}-1\right)^{1/2}.
\end{equation}
Replacing it into Eq.~\eqref{fe3} yields \begin{eqnarray}
	&&\frac{d \tilde m_\star}{d x_\star} = 8 x_\star^2 \tilde \beta_\star^2\left(
		\frac{\tilde E (1+\zeta)}{ \sqrt{A}} - 4\tilde \beta_\star^2
	\right),
\label{ge}
\end{eqnarray}
Since $\tilde \beta_\star = \sqrt{|\tilde \uplambda|}\tilde \beta$, then  the self-interacting spinor field constant coupling,  
$\tilde \uplambda$, is explicitly evinced  in Eq.~\eqref{ge}.
Therefore, Eq.~\eqref{ge} can be employed to derive the rescaled MGD Dirac star mass, 
$\tilde M\mapsto \tilde M_\star=\frac{Mm}{\sqrt{|\tilde \uplambda|}}{M_{\rm p}^2}$.
Numerical analysis yields the solutions of Eqs.~(\ref{ge0}, \ref{ge}),  given in Fig.~\ref{fi4},

In the general-relativistic, $\upsigma\to\infty$ limit, the  maximal  mass was derived in Ref. \cite{Dzhunushaliev:2019uft}, being $
	  0.4132 \sqrt{|\tilde \uplambda|} \frac{M_{\rm p}^2}{m},$ matching  Eq.~\eqref{41a}.

\section{Concluding remarks, discussion and outlook}
\label{4}
 MGD compact stellar distributions, coupled to self-interacting fermionic fields of spin-1/2, were scrutinized. The time-dependent spinor field solutions are asymptotically flat. MGD Dirac stars qualitatively resemble standard Dirac stars, as the MGD parameter is bounded, $
|l|\lesssim 6.1 \times 10^{-11}$, by  current experimental and observational data  \cite{Casadio:2015jva}, together with the  most strict bound for the brane tension $\upsigma \gtrsim  2.81\times10^6 \;{\rm MeV^4}$ \cite{Fernandes-Silva:2019fez}. 
Therefore, due to the small physical values of these two parameters that rule the deformation process in the MGD, the numerical solutions plotted in Figs. \ref{fi1} -- \ref{fi4}, for MGD Dirac stars, present qualitative profiles that are similar  
to the standard Dirac stars \cite{Dzhunushaliev:2018jhj}, as expected. The MGD-decoupling parameter $\zeta=0.1$ was adopted in all the numerical calculations. As discussed in Ref. \cite{Dzhunushaliev:2018jhj},  standard Dirac stars generated by linear spinors fields have tiny masses. Including non-linear spinor fields, in particular the self-interaction (\ref{selfii}), circumvents this feature. It thus makes possible to approach astrophysical MGD Dirac stars. The MGD Dirac star maximal mass increases as a function of the spinor self-interaction coupling constant, $|\tilde\uplambda|$, as illustrated in Fig. (\ref{fi2}). Moreover,  for the strictest  phenomenological bound for the brane tension $\upsigma \gtrsim  2.81\times10^6 \;{\rm MeV^4}$, the 
MGD Dirac star maximal mass was shown to increase in a lower rate, compared to higher order values of the brane tension.

The spinor self-interaction coupling constant, $\tilde\lambda$, in Eq. (\ref{selfii}), indicates stable compact self-gravitating configurations, whose maximal mass is given by Eq. (\ref{41a}).
This value of the mass, alternatively written as   $1.96(1+\zeta)\times 10^5 \sqrt{|\uplambda|}M_\odot ({\rm MeV^2}/m)$ corresponds to the mass of a MGD Dirac star that has similar order of magnitude as the Chandrasekhar mass, for fermions with mass $m\approx 1$ GeV. In the general-relativistic limit, when $\zeta\to0$ and $\upsigma\to0$, all the results in Ref. \cite{Dzhunushaliev:2018jhj} are recovered.

For MGD bosonic stellar distributions formed by Bose--Einstein condensates of gravitons, Ref. \cite{Casadio:2016aum}
showed that a critical local point of the star information entropy, as a function of the stellar configuration mass, indicates a transition between stability and instability against linear perturbations. 
The extended MGD case was discussed in Ref. \cite{Fernandes-Silva:2019fez}. In the case here studied, 
there is a family of MGD Dirac stars, parametrized by  central value of the spinor field $\beta_{\rm c}$ component. The mass of a MGD Dirac star can be expressed in terms of $\beta_{\rm c}$, existing a local critical mass for each value of $\beta_{\rm c}$, given by the solution of Eq. (\ref{ge}).
In addition to this analysis, the information entropy of MGD Dirac stars should take place to provide a final answer to 
their instability/stability conditions. The developments in Refs. 
\cite{Braga:2019jqg,Braga:2016wzx} can shed new light on this important problem, that is beyond the scope of this paper.

\subsection*{Acknowledgements}

RdR~is grateful to FAPESP (Grant No.  2017/18897-8) and to the National Council for Scientific and Technological Development  -- CNPq (Grants No. 303390/2019-0, No. 406134/2018-9 and No. 303293/2015-2), for partial financial support.

\bibliography{bib_DSS}

\end{document}